
\magnification=1200
\baselineskip=13pt
\overfullrule=0pt
\tolerance=100000

{\hfill \hbox{\vbox{\settabs 1\columns
\+ UR-1417 \cr
\+ ER-40685-865\cr
\+ hep-th/9504030\cr
}}}

\bigskip
\bigskip
\baselineskip=18pt

\centerline{\bf Properties of Nonlocal Charges in the Supersymmetric
Two Boson Hierarchy}

\vfill

\centerline{J. C. Brunelli}
\medskip
\centerline{and}
\medskip
\centerline{Ashok Das}
\medskip
\medskip
\centerline{Department of Physics and Astronomy}
\centerline{University of Rochester}
\centerline{Rochester, NY 14627, USA}
\vfill

\centerline{\bf {Abstract}}

\medskip
\medskip

We obtain the conserved, nonlocal charges for the supersymmetric two boson
hierarchy from fractional powers of its Lax operator. We show that these
charges
reduce to the ones of the supersymmetric KdV system under appropriate
reduction.
We study the algebra of the nonlocal, local and supersymmetry charges
with respect to the first and the second Hamiltonian structures of the system
and discuss how they close as a graded nonlinear cubic algebra.

\vfill
\eject

\bigskip
\noindent {\bf 1. {Introduction}}
\medskip

Integrable models have been studied extensively in the past [1-4]. The
supersymmetric generalizations of these models
have also  raised a lot of interest after the
introduction of the supersymmetric KdV equation (sKdV) [5,6] and the
supersymmetric nonlinear Schr\"odinger equations (sNLS) [7,8]. In a recent
paper [9] we have constructed the supersymmetric version of the Two Boson
equation
(sTB) [10-13] also known as the dispersive long water wave equation [14-16].
This
integrable system (the bosonic as well the supersymmetric version) has a very
rich structure since it is tri-Hamiltonian, has a nonstandard Lax
representation and reduces to  well known integrable systems under appropriate
reductions.

An interesting property of the supersymmetric integrable models is the
existence of nonlocal conservation laws. These conserved charges were first
obtained in [17] for the sKdV equation and in [7] for the sNLS equation through
group symmetry analysis of these equations. In [18] a nice interpretation for
the existence of the nonlocal charges for the sKdV was given: they can be
obtained as the Adler supertrace of odd powers of the fourth root of
the Lax operator.

In this paper we show that  nonlocal conserved charges also arise in
the sTB hierarchy and as
the supertrace of odd powers of the square root of the Lax operator of our
nonstandard system. The explicit form of the Hamiltonian structures of our
system allows us to calculate the algebra of these nonlocal charges,
the local ones and the supersymmetry  charge. We find the algebra to be a
nonlinear algebra with a cubic term much like the algebras of the
nonlocal charges of
the Heisenberg spin chain and the nonlinear sigma model [19-21].
This result
indicates the presence of some sort of a Yangian structure in the
supersymmetric
integrable models [22], a result which deserves further investigation.

Our paper is organized as follows. In sec. 2 we construct the first four local
conserved charges of the sTB equation (as well as the hierarchy)
and review its tri-Hamiltonian structure.
Here we also present the correct second and third Hamiltonian
structures of the system
which are slightly different from the ones given in [9].
In sec. 3 we construct the first three nonlocal conserved charges of
our model and discuss various properties of these charges.
In sec. 4 we calculate the algebra of the local, nonlocal and
supersymmetry charges, using both the first and the second
Hamiltonian structures and show that they satisfy a graded cubic algebra.
We present our conclusions as well as a short discussion on the algebra
in sec. 5.

\bigskip
\noindent {\bf 2. {Local Charges for the Supersymmetric Two Boson Hierarchy}}
\medskip

The supersymmetric Two Boson equation [9] is an integrable system represented
by a Lax operator of the form
$$
L = D^2 - (D \Phi_0) + D^{-1} \Phi_1\eqno(1)
$$
and a nonstandard Lax equation
$$
{\partial  L \over \partial  t} = \left[ L, (L^2)_{\geq 1}
\right]\eqno(2)
$$
which leads to the equations
$$
\eqalign{{\partial  \Phi_0 \over \partial  t} &=
 - (D^4 \Phi_0) + (D(D\Phi_0)^2)
                     + 2(D^2 \Phi_1)\cr
\noalign{\vskip 4pt}%
{\partial  \Phi_1 \over \partial  t} &=
 (D^4 \Phi_1) + 2 D^2((D \Phi_0) \Phi_1)\cr}\eqno(3)
$$
Here
$$
\eqalign{\Phi_0 &= \psi_0 + \theta J_0\cr
\Phi_1 &= \psi_1 + \theta J_1 \cr}\eqno(4)
$$
are two fermionic superfields, the covariant derivative in the superspace has
the form
$$
D = {\partial  \over \partial  \theta} + \theta {\partial
 \over \partial  x}\eqno(5)
$$
and the canonical dimensions of various quantities are  given by
$$
\displaylines{
\hfill
\eqalign{
[x]&=-1\cr
\noalign{\vskip 4pt}%
[t]&=-2\cr
\noalign{\vskip 4pt}%
[\theta] &= - {1 \over 2}\cr}
\hfill
\eqalign{
[\Phi_0] &= {1 \over 2}\cr
\noalign{\vskip 4pt}%
[\Phi_1] &= {3 \over 2}\cr
\noalign{\vskip 4pt}%
\phantom{[\Phi_1]} &\phantom{=} \cr}
\hfill(6)}
$$

The local conserved charges can be obtained from
$$
Q_n=\hbox{sTr}\,L^n=\int dz\,\hbox{sRes}\,L^n\qquad n=1,2,\dots \eqno(7)
$$
where ``sRes'' stands for the super residue which is defined to be the
coefficient of the $D^{-1}$ term in the pseudo super-differential operator
with $D^{-1}$ at the right
($D^{-1}=\partial^{-1}D$) and $z\equiv(x,\theta)$ is
the superspace coordinate. The first ones have the explicit form
\eject
$$
\eqalign{
Q_1=&-\int dz\, \Phi_1\cr
Q_2=&2\int dz\, (D\Phi_0)\Phi_1\cr
Q_3=&3\int dz\,\Bigl[(D^3\Phi_0)-(D\Phi_1)-(D\Phi_0)^2\Bigr]\Phi_1\cr
Q_4=&2\int
dz\,\Bigl[2(D^5\Phi_0)+2(D\Phi_0)^3+6(D\Phi_0)(D\Phi_1)-
3\left(D^2(D\Phi_0)^2\right)
\Bigr]\Phi_1\cr
}\eqno(8)
$$
Note that $[Q_n]=n$, they are bosonic and manifestly supersymmetric since
supersymmetry can be thought of as translations of the Grassmann coordinate
in the superspace (See also eq. (32).).

Defining the Hamiltonians as
$$
H_n={(-1)^{n+1}\over n}Q_n\eqno(9)
$$
we note that the system, eq.(3), has a tri-Hamiltonian structure of the form
$$
\partial_t\pmatrix{\Phi_0\cr
\noalign{\vskip 10pt}%
\Phi_1}
={\cal D}_1
\pmatrix{{\delta H_{3}\over\delta\Phi_0}\cr
\noalign{\vskip 10pt}%
{\delta H_{3}\over\delta\Phi_1}}=
{\cal D}_2
\pmatrix{{\delta H_{2}\over\delta\Phi_0}\cr
\noalign{\vskip 10pt}%
{\delta H_{2}\over\delta\Phi_1}}=
{\cal D}_3
\pmatrix{{\delta H_{1}\over\delta\Phi_0}\cr
\noalign{\vskip 10pt}%
{\delta H_{1}\over\delta\Phi_1}}
\eqno(10)
$$
where
$$
{\cal D}_1=\pmatrix{0 & -D\cr
\noalign{\vskip 5pt}%
-D & 0}\eqno(11)
$$
which yields the following nonvanishing
Poisson brackets in components (We list them for
our later use in the calculation of the charge algebra in section 4.)
$$
\eqalign{
\{\psi_0(x),\psi_1(y)\}_1=&-\delta(x-y)\cr
\noalign{\vskip 3pt}%
\{J_0(x),J_1(y)\}_1=&\delta'(x-y)\cr
}\eqno(12)
$$
The second structure is given by
$$
{\cal D}_2=\pmatrix{-2D-2D^{-1}\Phi_1D^{-1}+D^{-1}(D^2\Phi_0)D^{-1}&
D^3-D(D\Phi_0)+D^{-1}\Phi_1D\cr
\noalign{\vskip 20pt}%
-D^3-(D\Phi_0)D-D\Phi_1D^{-1}&-\Phi_1D^2-D^2\Phi_1}\eqno(13)
$$
which gives the nonvanishing  Poisson brackets in the components of the form
\eject
$$
\eqalign{
\{\psi_0(x),\psi_0(y)\}_2=&
\Bigl(\partial^{-1}J_0'\bigl(\partial^{-1}\delta(x-y)\bigr)\Bigr)
-2\Bigl(\partial^{-1}J_1\bigl(\partial^{-1}\delta(x-y)\bigr)\Bigr)
-2\delta(x-y)\cr
\noalign{\vskip 3pt}%
\{\psi_0(x),J_0(y)\}_2=
&\Bigl(\partial^{-1}\psi'_0\delta(x-y)\Bigr)
-2\Bigl(\partial^{-1}\psi_1\delta(x-y)\Bigr)\cr
\noalign{\vskip 3pt}%
\{J_0(x),J_0(y)\}_2=&2\delta'(x-y)\cr
\noalign{\vskip 3pt}%
\{\psi_0(x),\psi_1(y)\}_2=&
\Bigl(\partial^{-1}J_1\delta(x-y)\Bigr)-J_0\delta(x-y)+\delta'(x-y)\cr
\noalign{\vskip 3pt}%
\{\psi_0(x),J_1(y)\}_2=
&\psi'_0\delta(x-y)+\Bigl(\partial^{-1}\psi_1\delta'(x-y)\Bigr)\cr
\noalign{\vskip 3pt}%
\{J_0(x),\psi_1(y)\}_2=&\psi_1\delta(x-y)\cr
\noalign{\vskip 3pt}%
\{J_0(x),J_1(y)\}_2=&(J_0\delta(x-y))'-\delta''(x-y)\cr
\noalign{\vskip 3pt}%
\{\psi_1(x),J_1(y)\}_2=&2\psi_1\delta'(x-y)+\psi_1'\delta(x-y)\cr
\noalign{\vskip 3pt}%
\{J_1(x),J_1(y)\}_2=&J_1'\delta(x-y)+2J_1\delta'(x-y)\cr
}\eqno(14)
$$
In (14) as well as in (12) we remember that $\{A,B\}=-(-1)^{|A||B|}\{B,A\}$ and
the parenthesis limit the action of the inverse (integral) operators. We would
use this convention through out the paper. Introducing the recursion operator
$$
R={\cal D}_2{\cal D}_1^{-1}\eqno(15)
$$
the third Hamiltonian structure can be written as (One can write it out
explicitly, but its form is not very illuminating.)
$$
{\cal D}_3=R\,{\cal D}_2\eqno(16)
$$

Finally, through the recursion operator defined in (15) we can relate the local
conserved charges in a recursive manner as
$$
\pmatrix{{\delta H_{n+1}\over\delta\Phi_0}\cr
\noalign{\vskip 10pt}%
{\delta H_{n+1}\over\delta\Phi_1}}=R^\dagger
\pmatrix{{\delta H_{n}\over\delta\Phi_0}\cr
\noalign{\vskip 10pt}%
{\delta H_{n}\over\delta\Phi_1}}\eqno(17)
$$
where
$$
R^\dagger=\pmatrix{D^2-D^{-1}(D^2\Phi_0)+(D\Phi_0)+\Phi_1D^{-1}&
2(D\Phi_1)-2\Phi_1D-D^{-1}(D^2\Phi_1)\cr
\noalign{\vskip 20pt}%
2+D^{-2}\Phi_1D^{-1}-D^{-2}(D^2\Phi_0)D^{-1}&
-D^2-D^{-2}(D\Phi_1)+(D\Phi_0)+D^{-1}\Phi_1
}\eqno(18)
$$

Here we note that the second Hamiltonian structure in equation (13) differs
slightly from ${\widetilde{\cal D}}_2$ given in [9]. Even though
${\widetilde{\cal D}}_2$ also gives the right equations when used in (10)
and defines a recursion operator (as in (15)) which relates the lower order
conserved charges (as in (17)), it differs from
${\cal D}_2$ by the presence of nonlocal field dependent terms in the latter.
The difference in the two structures, therefore, gives no contribution to
the Hamiltonian equations (as can be explicitly checked) and yet is crucial for
Jacobi identity to hold.
In fact, using the prolongation
methods described in [23] (and generalized to
the supersymmetric systems in [24]), the bivector associated with
${\widetilde{\cal D}}_2$
$$
\Theta_{{\widetilde{\cal D}}_2}={1\over 2}\sum_{\alpha,\beta}\int dz\,
\left(({\widetilde{\cal D}}_2)_{\alpha\beta}\Omega_\beta\right)
\wedge\Omega_\alpha\qquad
\alpha,\beta=0,1\eqno(19)
$$
yields a field independent term under prolongation
$$
\hbox{\bf pr}\,{\vec v}_{{\widetilde{\cal D}}_2{\vec\Omega}}
(\Theta_{{\widetilde{\cal D}}_2})=
-\int dz\,(D\Omega_0)\wedge(D\Omega_0)\wedge(D\Omega_1) \neq 0
\eqno(20)
$$
which implies that ${\widetilde{\cal D}}_2$ does
not satisfy the Jacobi identity.
However, it can be checked, in a straightforward but tedious manner that
$$
\hbox{\bf pr}\,{\vec v}_{{\cal D}_2{\vec\Omega}}
(\Theta_{{\cal D}_2}) = 0\eqno(21)
$$
Therefore, ${\cal D}_2$  satisfies the Jacobi identity and represents a true
Hamiltonian operator. Similar results also hold for ${\cal D}_3$. Thus,
${\cal D}_1$, ${\cal D}_2$ and ${\cal D}_3$ represent the true Hamiltonian
structures of this theory.

\bigskip
\noindent {\bf 3. {Nonlocal Charges for the Supersymmetric Two Boson
Hierarchy}}
\medskip

Existence of nonlocal conserved charges in supersymmetric integrable models
such as the sKdV equation
were first recognized in [17]. However, in ref. [18], a systematic procedure
for their construction was given within the framework of
the Gelfand-Dikii formalism.
The authors in [18] realized that while the local charges for the sKdV can
be obtained from odd powers of the  square root of the Lax operator,
$L^{2n-1\over2}$, the nonlocal ones
can be obtained from odd powers of the quartic roots, $L^{2n-1\over4}$.
For the sTB, the situation is
quite similar. Since the local charges $Q_n$'s are
given by integer powers of $L$ (Eq.
(7)), we expect to generate conserved nonlocal charges $F_n$'s
from odd powers of the square root of $L$, that is,
$$
F_{2n-1\over2}=\hbox{sTr}\,L^{2n-1\over2}\qquad n=1,2,\dots\eqno(22)
$$
The square root of $L$ given in (1) can be written as
$$
L^{1/2}=D+a_0+a_1D^{-1}+a_2D^{-2}+a_3D^{-3}+a_4D^{-4}+a_5D^{-5}+\cdots
\eqno(23)
$$
where
$$
\eqalign{
a_0=&2(D^{-2}\Phi_1)-\Phi_0\cr
a_1=&-(D^{-1}\Phi_1)\cr
a_2=&(D^{-1}(\Phi_0\Phi_1))-2(D^{-2}((D\Phi_0)\Phi_1))
+\Phi_0(D^{-1}\Phi_1)-\Phi_1\cr
a_3=&{1\over2}(D^{-1}\Phi_1)^2-(D\Phi_0)(D^{-1}\Phi_1)+
(D^{-1}((D\Phi_0)\Phi_1))+(D\Phi_1)\cr
\int dz\,a_5=&\int dz\,\biggl[{\cal O}
(Da_1)-\Bigl(D^{-1}\bigl(2a_3(D^2a_0)-a_1a_3+(Da_1)(D{\cal O})\bigr)\Bigr)
\biggr]}\eqno(24)
$$
and we have defined (We note here that for the first three nonlocal charges
that we will list below, we do not need $a_4$ and need only the integrated
form of $a_5$.)
$$
{\cal O}\equiv (D^2a_0)+(Da_1)-2a_2\eqno(25)
$$
The grading of the coefficients are
$$
|a_n|=n+1\eqno(26)
$$
In what follows, the relation
$$
(-1)^{|A|}\bigl(D^{-1}(AB)\bigr)=A(D^{-1}B)-
\Bigl(D^{-1}\bigl((DA)(D^{-1}B)\bigr)\Bigr)\eqno(27)
$$
is very useful and can be easily proved through the Leibnitz rule. Also, to
perform integration by parts we need the generalized formula [18] which holds
for local functions $A$ and $B$,
$$
\int dz\,(D^nA)B=(-1)^{n|A|+{n(n+1)\over2}}\int dz\, A(D^nB),\quad\hbox{for
all }n\eqno(28)
$$
We note, however, that for nonlocal functions, the surface terms can not always
be neglected.

The first three nonlocal charges can be obtained after some long, but
straightforward, calculations
$$
\eqalign{
F_{1/2}=&-\int dz\, (D^{-1}\Phi_1)\cr
F_{3/2}=&-\int dz\,\biggl[{3\over2}(D^{-1}\Phi_1)^2-\Phi_0\Phi_1-
\Bigl(D^{-1}\bigl((D\Phi_0)\Phi_1\bigl)\Bigr)\biggr]\cr
F_{5/2}=&-\int dz\,\biggl[\,{1\over6}(D^{-1}\Phi_1)^3-
\bigl(5(D^{-2}\Phi_1)\Phi_1
-2\Phi_0\Phi_1-3(D\Phi_1)-(D^{-1}\Phi_1)^2\bigr)(D\Phi_0)\cr
&\phantom{2\int dz\,\Bigl[}
+\Bigl(D^{-1}\bigl((D\Phi_1)\Phi_1+\Phi_1(D\Phi_0)^2-
(D\Phi_1)(D^2\Phi_0)\bigr)\Bigr)\biggr]\cr
}\eqno(29)
$$
These charges can be explicitly checked to be conserved under the flow (3). In
fact, it can be shown from the structure of the Lax equation (2) that the
nonlocal charges defined in (22) are indeed conserved under the flow of
the sTB hierarchy.

It is worth noting here some of the interesting properties of these charges
before evaluating their algebra. First of all, all the nonlocal charges are
fermionic and $[F_{2n-1\over2}]={2n-1\over2}$. Second, even though these
charges are expressed as superspace integrals, they are not invariant under the
supersymmetry transformations of the system (See also eq. (34).). This is
mainly
because of the nonlocality in the integrands. (We emphasize here that a
superspace integral of a local function of superfields is automatically
supersymmetric -- not necessarily true for nonlocal functions.) However, we
also note that there is nothing wrong with these charges not being
supersymmetric. After all, the supersymmetry charge, in these integrable
models, is not supersymmetric. Rather it satisfies the graded Lie algebra
$$
\{Q,Q\} = P\eqno(30)
$$
where $P$ denotes the momentum operator.
As we will show, these nonlocal charges are also part of an interesting graded
algebra.

We also note that the nonlocal charges of the sTB hierarchy in (29) reduce to
those of the sKdV hierarchy, up to normalizations [18],
when we set $\Phi_0=0$. This is not surprising
since we have already shown earlier [9] that the sKdV system is contained in
the
sTB system. However, unlike the sKdV system, here  the nonlocal charges are
not related recursively by either $R$ or $R^\dagger$ of eq. (18).
In fact, we have tried to
construct systematically an  operator which will relate the nonlocal charges
recursively, but have not succeeded. We can only speculate that these fermionic
charges presumably generate fermionic flows with distinct Hamiltonian
structures of their own
which in turn can give a ``recursion'' operator connecting them.
However, we would also like to add that these fermionic flows cannot have a Lax
representation of the form
${\partial L\over\partial\beta}=[L,(L^{1/2})_{\geq1}]$, with $\beta$ as
the odd time, since such an equation is not  consistent.
In [25] odd flows based on Jacobian supersymmetric
KP-hierarchies were studied for the sKdV case. This  may give some insight to
our present system if we use, instead, a nonstandard supersymmetric
KP-hierarchy [26]. We do not have anything to add to this at this time.
Finally, using the transformation given in [9] it is possible
to obtain the nonlocal charges for the sNLS equation which should coincide
with the ones obtained in [7].
\bigskip
\noindent {\bf 4. {Algebra of the Charges}}
\medskip

First, let us note that the sTB equation is invariant under
supersymmetry transformations generated by the following conserved
fermionic charge (As we have noted earlier, the supersymmetry charge is not
supersymmetric and, consequently, cannot be written as an integral of a local
function in superspace. Therefore, it is much more convenient to work in
components.)
$$
Q=-\int dx\,\left(\psi_1J_0+\psi_0J_1\right)\eqno(31)
$$
This charge together with  the first Hamiltonian structure in (12), generates
the supersymmetry transformations
$$
\eqalign{
\delta J_0=&\epsilon\{Q,J_0\}_1=\epsilon\psi_0'\cr
\delta J_1=&\epsilon\{Q,J_1\}_1=\epsilon\psi_1'\cr
\delta\psi_0=&\epsilon\{Q,\psi_0\}_1=\epsilon J_0\cr
\delta\psi_1=&\epsilon\{Q,\psi_1\}_1=\epsilon J_1\cr
}\eqno(32)
$$
where $\epsilon$ is a constant Grassmann parameter of transformation.

It can now be easily shown with eqs. (31) and (12) that
$$
\{Q,Q\}_1=-Q_2\eqno(33)
$$
which implies that the supersymmetry charge is not supersymmetric --
rather it satisfies a graded Lie algebra. (We note here that, for
conventional Lie algebra symmetries, the change in any variable $A$ under a
symmetry transformation generated by a charge $G$ is given, except for factors,
as $\delta A = \{G,A\}$. The above result, therefore, would say that under a
supersymmetry transformation $\delta Q\neq 0$.).

As we have pointed out
earlier even the nonlocal charges are not invariant under supersymmetry
transformations and, in fact, we can easily calculate and see that
$$
\eqalign{
\{Q,F_{1/2}\}_1=&Q_1 \cr
\{Q,F_{3/2}\}_1=&{1\over2}Q_2 \cr
\{Q,F_{5/2}\}_1=&{1\over3}Q_3+{1\over24}Q_1^3 \cr
}\eqno(34)
$$
We also note that since the supersymmetry  charge $Q$ as well as the nonlocal
charges $F_{2n-1\over2}$ are conserved, they are in involution with the local
charges $Q_n$ (This can be explicitly checked for the flow in (3).), i.e,
$$
\eqalign{
\{Q_n,Q_m\}_1=&0\cr
\{Q_n,F_{2m-1\over2}\}_1=&0\cr
\{Q_n,Q\}_1=&0\cr
}\eqno(35)
$$
The algebra of the  nonlocal charges is interesting as well. We list here only
the first few relations
$$
\eqalign{
\{F_{1/2},F_{1/2}\}_1=&0\cr
\{F_{1/2},F_{3/2}\}_1=&Q_1 \cr
\{F_{1/2},F_{5/2}\}_1=&Q_2 \cr
\{F_{3/2},F_{3/2}\}_1=&2Q_2 \cr
\{F_{3/2},F_{5/2}\}_1=&{7\over3}Q_3+{7\over24}Q_1^3 \cr
\{F_{5/2},F_{5/2}\}_1=&3Q_4-{5\over 8}Q_1^2 Q_2 \cr
}\eqno(36)
$$
This  shows that the algebra of the
conserved charges $Q$, $Q_n$ and $F_{2n-1\over2}$, at least at these orders,
closes with respect to the
first Hamiltonian structure in (12). However, the algebra is not a linear Lie
algebra. Rather it is a graded nonlinear algebra where the nonlinearity
manifests in a cubic term. In fact, the canonical dimensions of the charges
allows for higher order nonlinearity to be present even in the algebraic
relations of these lower order charges, but
the algebra appears to involve only a cubic nonlinearity.
The Jacobi identity is seen to be trivially satisfied for this algebra since
the $Q_n$'s are in involution with all the fermionic charges ($Q_n$'s are the
Hamiltonians for the system and the fermionic charges are conserved.).
We note here that the cubic terms in (34) and (36)
arise from boundary contributions when nonlocal terms are involved. This can be
seen simply using the following example. If we use
the following realization of the inverse  operator
$$
\bigl(\partial^{-1} f(x)\bigr) ={1\over 2}\int d y \, \epsilon(x-y)f(y)
\,,\qquad \epsilon(x)= \cases{-1,\,& $x<0$\cr \phantom{-}0,\, &$x=0$\cr
+1,\, &$x>0$\cr}
\quad \eqno(37)
$$
then, this will lead, for instance, to boundary terms of the type
$$
\int dz\,(D^{-1}\Phi_1)^2\Phi_1={1\over3}\int dz\,D(D^{-1}\Phi_1)^3=
-{1\over12}Q_1^3\eqno(38)
$$
and this is the origin of the nonlinear terms. We would also like to point out
that we have not succeeded in finding a closed form expression for a general
element of the algebra in (34) and (36), but we will discuss in the conclusion
why we believe the algebra to be a cubic algebra.

The sTB hierarchy has three distinct Hamiltonian structures. Consequently,
one can ask  other interesting questions such as what transformations does $Q$
generate with the second structure or what is the algebra of the charges with
respect to the second structure and so on. The supersymmetry transformations
generated by $Q$ with respect to the second structure can be easily calculated
and have the form
$$
\eqalign{
\delta J_0=&\epsilon\{Q,J_0\}_2=\epsilon\left(2\psi_1'+J_0\psi_1+(\psi_0J_0)'-
\psi_0''+(\partial^{-1}J_1)\psi_0'-2(\partial^{-1}J_1)\psi_1\right)\cr
\delta J_1=&\epsilon\{Q,J_1\}_2=\epsilon
\left(\psi_1''+2(J_0\psi_1)'+(J_1\psi_0)'
-(\psi_1(\partial^{-1}J_1))'\right)\cr
\delta\psi_0=&\epsilon\{Q,\psi_0\}_2=\epsilon
\left(J_0^2-J_0'-J_0(\partial^{-1}J_1)+
\psi_0\psi_0'+2J_1+2(\partial^{-1}((\partial^{-1}J_1)J_1))\right)\cr
\delta\psi_1=&\epsilon\{Q,\psi_1\}_2=\epsilon
\left(2\psi_0'\psi_1+\psi_0\psi'_1+J_1'+
J_1(\partial^{-1}J_1)+J_1J_0\right)\cr
}\eqno(39)
$$
We note that these are highly nonlinear fermionic transformations and are
nonlocal. We have explicitly verified
that these transformations define a symmetry of
the equations of motion of the supersymmetric Two Boson system.

The algebra of the charges can again be calculated with respect to the second
Hamiltonian structure and we obtain
$$
\{Q,Q\}_2={2\over3}Q_3-{1\over6}Q_1^3\eqno(40)
$$
This shows that, with respect to the second structure, $Q$ satisfies a cubic
graded algebra.  Furthermore, with  the nonlocal charges, it gives
$$
\eqalign{
\{Q,F_{1/2}\}_2=&-{1\over2}Q_2 \cr
\{Q,F_{3/2}\}_2=&-{1\over3}Q_3+{1\over12}Q_1^3 \cr
\{Q,F_{5/2}\}_2=&-{1\over4}Q_4+{1\over16}Q_1^2Q_2 \cr
}\eqno(41)
$$
\noindent The charges $Q$ and $F_{2n-1\over2}$ are in involution with $Q_n$
$$
\eqalign{
\{Q_n,Q_m\}_2=&0\cr
\{Q_n,F_{2m-1\over2}\}_2=&0\cr
\{Q_n,Q\}_2=&0\cr
}\eqno(42)
$$
and the algebra among the  nonlocal charges has the form
$$
\eqalign{
\{F_{1/2},F_{1/2}\}_2=&0\cr
\{F_{1/2},F_{3/2}\}_2=&-{1\over2}Q_2 \cr
\{F_{1/2},F_{5/2}\}_2=&-{2\over3}Q_3-{1\over12}Q_1^3 \cr
}\eqno(43)
$$
We see that the charges satisfy a graded algebra with respect to the second
structure and the algebra continues to be a cubic algebra. In fact, we note
that (41) and (43) represent a sort of  shifting of (34) and (36). We believe
that these general qualitative features would continue to hold even for
the third Hamiltonian structure.

\bigskip
\noindent {\bf 5. {Conclusion}}
\medskip

In this letter, we have constructed the conserved, nonlocal, fermionic charges
for the sTB hierarchy. We have also presented the three, correct Hamiltonian
structures of the system satisfying Jacobi identity which is
checked through super-prolongation methods.
Our nonlocal charges are not related by the recursion
operator of the theory even though they reduce to the charges of the sKdV
system
under appropriate reduction. The fermionic, nonlocal charges are not
supersymmetric. Rather, the bosonic and the fermionic charges of the system
$Q_n$, $Q$ and $F_{{2n-1}\over2}$
satisfy a graded algebra which is nonlinear and
appears to be cubic. This structure of our algebra continues to hold even when
evaluated with the second Hamiltonian structure of the system and we believe
the third structure also will lead to similar conclusion.

We would like to point out here that the appearance of cubic terms in the
algebra of nonlocal charges in this integrable model is nothing new. When
carefully evaluated, such terms are also present in the algebra of nonlocal
charges in the case of the sKdV system [18] even though it has not been
observed
before. For example, if we take the sKdV equation,
$\Phi_t=-(D^6\Phi)+3D^2(\Phi(D\Phi))$ and use the nonlocal charge
$$
J_{3/2}=\int dz\,(D^{-1}\Phi)^2\eqno(44)
$$
as well as the second Hamiltonian structure, as in [18] (The first Hamiltonian
structure in the case of sKdV is not very convenient for calculations.),
we obtain (See eq. (38).)
$$
\{J_{3/2},J_{3/2}\}=4\int dz\,\left(\Phi(D\Phi)+\Phi(D^{-1}\Phi)^2\right)=
4H_3-{1\over3}H_1^3\eqno(45)
$$
where we have defined
the conserved local charges of sKdV as (the normalization is different from
[18])
$$
H_{2n-1}={2^{2n-1}\over 2n-1}\hbox{Tr}\,L^{2n-1\over2}\eqno(46)
$$

Cubic algebras have been found earlier in studies of other systems such as the
Heisenberg spin chains and the nonlinear sigma models [19-21] and appear to be
a
common feature when nonlocal charges are involved. In fact, on general grounds,
one can argue that the nonlinearity in these algebras can be high. However, it
is possible to redefine the generators in a highly nonlinear and nontrivial way
such that the algebra is indeed cubic. This is quite well known in the case of
the nonlinear sigma model [21]. Here we indicate how it can be done for the
sKdV system which is simpler compared to the sTB hierarchy.
The following algebra can be derived for the sKdV in a straightfoward manner
$$
\eqalign{
\{J_{1/2},J_{1/2}\}=&-H_1\cr
\{J_{1/2},J_{3/2}\}=&0\cr
\{J_{1/2},J_{5/2}\}=&-6H_3-{1\over4}H_1^3\cr
\{J_{3/2},J_{3/2}\}=&4H_3-{1\over3}H_1^3\cr
\{J_{3/2},J_{5/2}\}=&0\cr
\{J_{5/2},J_{5/2}\}=&-36H_5-{9\over80}H_1^5\cr
}\eqno(47)
$$
Here the structure constants can be further simplified by rescaling
$J_{2n-1\over2}$, but we do not bother about it here.
The important point to note
is the appearance of the quintic term in the bracket of $J_{5/2}$ with itself.
Let us note, however, that we can redefine
$$
\eqalign{
{\hat J}_{1/2}=&J_{1/2}\cr
{\hat J}_{3/2}=&J_{3/2}\cr
{\hat J}_{5/2}=&J_{5/2}+\alpha H_1^2 J_{1/2}\cr
}\eqno(48)
$$
where $\alpha$ can be chosen such that the algebra becomes cubic. (Choice of
$\alpha$ eliminates the quintic term. We note here that $\alpha$ turns out to
be complex in this case. This may suggest a different normalization for the
charges.) From this
we strongly believe that one can redefine the charges even in
the case of sTB such that the right hand side of the algebra in (34), (36),
(41) and (43) will have the closed form structure
$$
a\,{\hat Q}_n+b\!\!\!\sum_{p+q+\ell=n}\!\!\!
{\hat Q}_p {\hat Q}_q {\hat Q}_\ell\eqno(49)
$$
where $a$ and $b$ are numerical factors and $n$ is the sum of the canonical
dimensions of the left hand side of the algebra.

We would conclude by noting here that the cubic algebras of this sort can be
related with Yangians [19-22]. So, it is natural to expect
that the algebra, in the present systems (both sKdV and sTB) also corresponds
to a Yangian. There is, however, a difficulty with this. Namely, a Yangian
starts out with a non-Abelian Lie algebra for the local charges. Here, in
contrast, the algebra of the local charges, $Q_n$'s, is involutive (Abelian).
There may still be an underlying Yangian structure in this algebra and this
remains an open question.

\bigskip
\noindent {\bf Acknowledgements}
\medskip

This work was supported in part by the U.S. Department of Energy Grant No.
DE-FG-02-91ER40685. J.C.B. would like to thank CNPq, Brazil, for
financial support.

\vfill\eject

\noindent {\bf {References}}
\bigskip

\item{1.} L.D. Faddeev and L.A. Takhtajan, ``Hamiltonian Methods in
the Theory of Solitons'' (Springer, Berlin, 1987).

\item{2.} A. Das, ``Integrable Models'' (World Scientific, Singapore,
1989).

\item{3.} M.J. Ablowitz and P.A. Clarkson, ``Solitons, Nonlinear
Evolution Equations and Inverse Scattering'' (Cambridge, New York, 1991).

\item{4.} L. A. Dickey, ``Soliton Equations and Hamiltonian Systems'' (World
Scientific, Singapore, 1991).

\item{5.} Yu. I. Manin and A. O. Radul, Commun. Math. Phys. {\bf 98}, 65
(1985).

\item{6.} P. Mathieu, J. Math. Phys. {\bf 29}, 2499 (1988).

\item{7.} G.H.M. Roelofs and P.H.M. Kersten, J. Math. Phys. {\bf 33}, 2185
(1992).

\item{8.} J.C. Brunelli and A. Das, J. Math. Phys. {\bf 36}, 268 (1995).

\item{9.} J.C. Brunelli and A. Das, Phys. Lett. {\bf B337}, 303 (1994).

\item{10.} H. Aratyn, L.A. Ferreira, J.F. Gomes and A.H. Zimerman, Nucl.
 Phys. {\bf B402}, 85 (1993); H. Aratyn, L.A. Ferreira, J.F. Gomes and A.H.
Zimerman, ``Lectures at the VII J. A. Swieca Summer School'', January 1993,
hep-th/9304152; H. Aratyn, E. Nissimov and S. Pacheva, Phys. Lett. {\bf B314},
41 (1993).

\item{11.} L. Bonora and C.S. Xiong, Phys. Lett. {\bf B285}, 191 (1992); L.
Bonora and C.S. Xiong, In. J. Mod. Phys. {\bf A8}, 2973 (1993).

\item{12.} M. Freeman and P. West, Phys. Lett. {\bf 295B}, 59 (1992).

\item{13.} J. Schiff, \lq\lq The Nonlinear Schr\"odinger Equation and
Conserved Quantities in the Deformed Parafermion and SL(2,{\bf R})/U(1)
Coset Models", hep-th/9210029.

\item{14.} L.J.F. Broer, Appl. Sci. Res. {\bf 31}, 377 (1975).

\item{15.} D.J. Kaup, Progr. Theor. Phys. {\bf 54}, 396 (1975).

\item{16.} B.A. Kupershmidt, Commun. Math. Phys. {\bf 99}, 51 (1985).

\item{17.} P. H. M. Kersten, Phys. Lett. {\bf A134}, 25 (1988).

\item{18.} P. Dargis and P. Mathieu, Phys. Lett. {\bf A176}, 67 (1993).

\item{19.} D. Bernard and A. LeClair, Commun. Math. Phys. {\bf 142}, 99 (1989);
D. Bernard, ``An Introduction to Yangian Symmetries'', in Integrable
Quantum Field Theories, ed. L. Bonora et al., NATO ASI Series B: Physics vol.
310 (Plenum Press, New York, 1993), and references therein.

\item{20.} J. Barcelos-Neto, A. Das, J. Maharana, Z. Phys. {\bf 30C}, 401
(1986);

\item{21.} E. Abdalla, M. C. B. Abdalla, J. C. Brunelli and A. Zadra, Commun.
Math. Phys. {\bf 166}, 379 (1994), and references therein.

\item{22.} T. Curtright and C. Zachos, Nucl. Phys. {\bf B402}, 604 (1993).

\item{23.} P. J. Olver , ``Applications of Lie Groups to Differential
Equations'', Graduate Texts in Mathematics, Vol. 107 (Springer, New York,
1986).

\item{24.} P. Mathieu, Lett. Math. Phys. {\bf 16}, 199 (1988).

\item{25.} E. Ramos, Mod. Phys. Lett. {\bf A9}, 3235 (1994).

\item{26.} J. C. Brunelli and A. Das, ``A Nonstandard Supersymmetric KP
Hierarchy'', University of Rochester preprint UR-1367 (1994) (also
hep-th/9408049).

\end